%% file: main.tex
\documentclass[a4paper,12pt,titlepage]{article}
\usepackage[utf8]{inputenc}
\usepackage{amsmath}
\usepackage{amsfonts}
\usepackage{amssymb}
\usepackage{graphicx}
\usepackage{setspace}
\usepackage{adjustbox}
\usepackage{xcolor}
\usepackage{csquotes}
\usepackage{listings}
\usepackage{lscape} 
\usepackage{threeparttable}
\usepackage[normalem]{ulem}
\usepackage[backend=bibtex, style=authoryear, sorting=nyt]{biblatex}
\RequirePackage[rmargin=1in,lmargin=1in, lines=25]{geometry}
\usepackage{authblk}

\usepackage{subfiles} 

\title{A Bayesian semi-parametric approach for inference on the population partly conditional mean from longitudinal data with dropout}
\author[1,2,3,*]{Josefsson, Maria}

\author[4]{Daniels, Michael J.}

\author[3,5]{Pudas, Sara}

\affil[1]{Department of Statistics, USBE, Ume{\aa} University, Sweden.}

\affil[2]{Centre for Demographic \& Ageing Research, Ume{\aa} University, Sweden.}

\affil[3]{Ume{\aa} Center for Functional Brain Imaging, Ume{\aa} University, Sweden.}

\affil[4]{Department of Statistics, University of Florida, USA.}

\affil[5]{Department of Integrative Medical Biology, Ume{\aa} University, Sweden}

\affil[*]{Corresponding author, E-mail: maria.josefsson@umu.se}
\date{}

\begin{document}
\thispagestyle{plain}
\begin{center}
    {\large
    \textbf{A Bayesian semi-parametric approach for inference on the population partly conditional mean from longitudinal \\
    data with dropout}}
        
\vspace{0.25cm}
\textsc{Josefsson, Maria$^{1,2,3,*}$, Daniels, Michael J.$^4$, \& Pudas, Sara$^{3,5}$}
\end{center}

\begin{center}{\footnotesize
$^1$Department of Statistics, USBE, Ume{\aa} University, Sweden.

$^2$Centre for Demographic \& Ageing Research, Ume{\aa} University, Sweden.

$^3$Ume{\aa} Center for Functional Brain Imaging, Ume{\aa} University, Sweden.

$^4$Department of Statistics, University of Florida, USA.

$^5$Department of Integrative Medical Biology, Ume{\aa} University, Sweden.

$^*$Corresponding author, E-mail: maria.josefsson@umu.se}
\end{center}

\begin{center}
\textbf{Summary}
\end{center}
Studies of memory trajectories using longitudinal data often result in highly non-representative samples due to selective study enrollment and attrition. An additional bias comes from practice effects that result in improved or maintained performance due to familiarity with test content or context. These challenges may bias study findings and severely distort the ability to generalize to the target population. In this study we propose an approach for estimating the finite population mean of a longitudinal outcome conditioning on being alive at a specific time point. We develop a flexible Bayesian semi-parametric predictive estimator for population inference when longitudinal auxiliary information is known for the target population. We evaluate sensitivity of the results to untestable assumptions and further compare our approach to other methods used for population inference in a simulation study. The proposed approach is motivated by 15-year longitudinal data from the Betula longitudinal cohort study. We apply our approach to estimate lifespan trajectories in episodic memory, with the aim to generalize findings to a target population.


\begin{center}
\textsc{1 Introduction}
\end{center}
Studies of lifespan trajectories in memory using longitudinal data present numerous methodological challenges including highly non-representative samples, due to selective study enrollment and attrition, and practice effects, which results in improved or maintained performance due to familiarity with test  (\cite{weuve2015guidelines}). These challenges may bias study findings and severely distort the ability to generalize to the target population, even in well-designed studies.

Improvements in test scores upon repeated assessment due to practice effects (PEs)are well documented in the cognitive aging literature (e.g.  \cite{salthouse2010influence}).  PEs arise from increased familiarity with the assessment tools and may result in underestimating the  decline  in  memory  in  longitudinal  studies. Common statistical approaches for handling PEs include: ignoring, specifying an indicator for the first assessment, or modeling a linear trend (\cite{vivot2016jump}). These approaches however remain controversial due to, for example, the strong assumption that the practice gains are the same across the age range or the difficulties in separating effects of within-person change from PEs (\cite{weuve2015guidelines, hoffman2011pe}).

Attrition is inevitable especially if individuals in the studied population are followed over a long time period. Standard methods often rely on the assumption of missing at random (MAR) which is invalid if the missingness is missing not at random (MNAR) or due to death. One natural way of dealing with MNAR missingness in any analysis of incomplete data is to explore sensitivity to deviations from the MAR assumption. This can be accomplished in a Bayesian setting, by introducing sensitivity parameters that incorporate prior beliefs about the differences between respondents and non-respondents and assigning an appropriate prior distribution (\cite{daniels08}). However, attrition due to death must be treated differently if a participant's death during follow-up truncates the outcome process (\cite{kurland2009longitudinal}). For example, in research on cognitive aging, cognition after death is not defined and not of interest to study. 

Joint models accompanied with additional postprocessing have been proposed to address the combination of dropout and death (\cite{rizopoulos2012joint}). This so called mortal cohort inference truncate participants outcomes after death, either by conditioning on the sub-population who are still alive at that time point, i.e. partly conditional inference (Kurland 2005), or by conditioning on the subpopulation that would survive irrespective of exposure, i.e. principal stratification (\cite{frangakis02, frangakis07}). The latter approach has been proposed to handle truncation by death when interest is in estimating the survivors average causal effect (e.g. \cite{josefsson2016causal}, \cite{mcguinness2019comparison}, \cite{shardell2018joint}). In contrast, partly conditional inference has been proposed for estimating non-causal associations. For example by using augmented inverse probability weighting for longitudinal cohort data with truncation by death and MAR dropout (\cite{wen2018methods}) as well as MNAR dropout (\cite{wen2018semi}). In addition, \citeauthor{li2018accommodating} (\citeyear{li2018accommodating}) proposed an approach for semicompeting risk data with MNAR missingness and death. 
This type of inference can for example be of interest, when the studied outcome is related to the health of all individuals who are alive. For example, in studies of health promotion and/or disease prevention among elderly.
These methods, however, fails to account for the complex nature of the sampling design such as selective study enrollment. 

While death is not viewed as a source of bias in longitudinal studies, because the objective is to generalize to those who are alive, selection and non-response bias (and additionally practice effects) often are, and therefore need to be addressed. Post-stratification (PS) is a technique in survey analysis that uses auxiliary information on the finite population to adjust for known or expected discrepancies between the sample and the population. A classical weighting estimator for finite population inference is the Horvitz-Thompson (HT) estimator (\cite{horvitz1952generalization}). Although the HT estimator is design unbiased, it can potentially be very inefficient. In contrast, a model-based (MB) approach specifies a model for the study outcome, usually a regression model, which is then used to make predictions for the population, and hence, finite population quantities. Predictions are calculated by plugging in the auxiliary variables for all units in the population in the working model. Model-based inferences generally will match or outperform design-based approaches if the model is correctly specified (\cite{little2004model}).
However, correct model specification can be difficult when there is a large set of regressors, the relationship is non-linear and/or includes interaction terms, and there are multiple observation times. 

Several approaches for poststratification adjustment in a cross-sectional setting have shown improved performance compared to the HT and the MB regression estimator. Multilevel regression combined with poststratification (MRP; e.g. \cite{gelman1997poststratification}, \cite{park2004bayesian}) have shown improved performance compared to standard weighting and MB regression estimators, and was used for producing accurate population estimates from a nonrepresentative sample (\cite{wang2015forecasting}).
The general regression estimator (GREG; \cite{deville1992calibration}) is a dual-modelling strategy that combines prediction and weighting. The approach require both a model for the outcome and the participation mechanism, and is double robust in the sense that it remains consistent if either one of the models are correctly specified. \citeauthor{bisbee2019barp} (\citeyear{bisbee2019barp}) combined a semi-parametric machine learning approach (Bayesian Additive Regression Trees (BART); \cite{chipman2010bart}) with post-stratification for predicting opinions using cross-sectional data. \citeauthor{kern2016assessing} (\citeyear{kern2016assessing}) showed in a simulation study that BART and inverse probability weighting using random forests performed better than approximately doubly robust estimators for estimating the target population average treatment effect. Although showing improved performance compared to MRP and GREG, previous semi-parametric approaches for population inference are \textit{not valid for longitudinal data with dropout and deaths}. 
 
In this study we propose an approach for estimating the finite population mean of a longitudinal continuous outcome conditioning on being alive at a specific time point, i.e. the population partly conditional mean (PPCM). Specifically, we develop a flexible Bayesian semi-parametric predictive estimator, when longitudinal auxiliary information is known for all units in the target population. The approach is to specify observed data models using BART and then to use assumptions with embedded sensitivity parameters to identify and estimate the PPCM. We evaluate sensitivity of the results to untestable assumptions on MNAR dropout and PEs, and further compare our approach to other methods used for population inference in a simulation study.

We are motivated by the Betula study, a prospective cohort study on memory, health and aging. The aim of the current paper is to extend previous results on cognitive lifespan trajectories (e.g. \cite{ronnlund05}, \cite{gorbach2017longitudinal}) by considering population partly conditional inference with MNAR missingness and practice effects. By using longitudinal micro-data from Statistics Sweden and the National Board of Health and Welfare, for both the sample and the target population, we are able to adjust for potential discrepancies in auxiliary variables and thereby improve generalizability of study findings.

The remainder of the paper is organized as follows. In Section 2, we present a motivating example. In Section 3 we present a MB approach for estimating the PPCM using longitudinal data with dropout and deaths and in Section 4 we describe a Bayesian semi-parametric modeling approach. In Section 5 we provide results from a simulation study and in Section 6 results from the empirical example using the Betula data. Conclusions are given in Section 7.


\begin{center}
\textsc{2 Motivating example}
\end{center}
The aim of the empirical study is to estimate lifespan trajectories in episodic memory, with the goal to generalize findings to a target population. Two separate sources of data are available for this study; a longitudinal cohort study and a longitudinal database covering the target population. 

The Betula study is a population-based cohort study with the objective to study how memory functions change over time and to identify risk factors for dementia (\cite{nilsson97}). The participants were randomly recruited, stratified by age, from the population registry in the Umeå municipality of Sweden. We consider longitudinal data from the first sample (S1) and four waves of data collection (T1-T4). There was five years in between each wave and the first wave of data collection was initiated in 1988-1990. A total of n=1000 participants were included, 100 participants from each of the 10  age-cohorts: $35, 40, \ldots , 80$. In order to obtain a total of 100 subjects in each of the 10 different cohorts, 1,976 persons had to be contacted. Of the 976 that never entered the study, 259 could not be reached, 130 had a illness (including dementia) to the extent that they could not participate, and 481 declined to participate (Nilsson et al., 1997). Memory was assessed at each wave using a composite of five episodic memory tasks, range: 0 - 76, where a higher score indicate better memory (\cite{josefsson12}).

The second data source is the Linneaus Database (\cite{malmberg2010longitudinal}), a longitudinal database covering every Swedish resident. The database includes annual data from Statistics Sweden and the National Board of Health and Welfare (similar information as for the Betula sample). In this study we consider micro data for every unit of the population in the Umeå municipality who were alive and non-demented in 1990 as the target population ($N = 9203$). Although longitudinal data is available annually we restrict data to the years: 1990, 1995, 2000, and 2005, approximately corresponding to the years of testing in the Betula study. 

A set of continuous and categorical auxiliary variables, linked to both selective study enrolment and memory, are included in both data sources. From the cause of death register we know death year for each deceased individual. In the Betula sample 29.1\% died during the study period and 20.0\% of the target population, and there were 128 and 806 dementia cases (12.8\% and 8.8\%) in the sample and population respectively. 

\begin{center}
\textsc{3 A model-based approach for population inference}

\textit{3.1 Cross-sectional setting}
\end{center}
Consider a finite population $U=\{ 1,2,\ldots , N \}$. For each individual $i \in U$, a $[1 \times K]$ vector of auxiliary information $\mathbf{x}_i$ is observed. A probability sample $c$ of size $n$ is drawn from $U$. 
Let $y_i$ be the continuous study variable which is observed if $i \in c$, and $p(y_i \mid \cdot)$ refers to a probability density function. Suppose interest is in studying the finite population mean, $\mu_U = N^{-1} \sum_{i \in U} y_i$. If we assume the participation mechanism is ignorable in the sense that $y_i$ and $i \in c$ are conditionally independent given $\mathbf{x}_i$ (\cite{rubin1976inference}), the joint distribution of the outcome, participation mechanism and auxiliary information can be factored into three conditional distributions $\prod_{i \in U} p(y_i \mid \mathbf{x}_i) p(i \in c \mid \mathbf{x}_i) p(\mathbf{x}_i)$.

A model-based (MB) prediction estimator incorporates the relationship between $\mathbf{x}$ and $y$ into the estimation of the population mean by $m(\mathbf{x}_i)=E(Y_i \mid X_i=x_i)$ and considers the finite population values $\{y_i ; i \in U\}$ to be realizations of the model. Conditional on the auxiliary variables $\mathbf{x}_i$, let $p(y_i \mid \mathbf{x}_i)$ be a model of the form $ y_i = m(\mathbf{x}_i) + \varepsilon_i $, where $\varepsilon_i$ are uncorrelated and $E( \varepsilon_i )=0$ and $Var( \varepsilon_i )=\sigma^2$. The predictions for each individual in the population, $\hat{y}_i$, are calculated by plugging in their auxiliary information $\mathbf{x}_i$ for all $i \in U$ in the working model $m(\mathbf{x}_i)$. A MB estimator of the population mean is obtained by 
$\hat{\mu}^{MB} = \frac{1}{N} \sum_{i \in U} \hat{m}(\mathbf{x}_i)$, where $\hat{m}(\mathbf{x}_i)$ is the estimated mean function.
We consider a setting with two separate data sources. Thus, we can not separate out the sample participants in the data covering the target population. Predictions must therefore be made for all participants in the target population.

\begin{center}
    \textit{3.2 Longitudinal data with dropout and death}
\end{center}
We now consider the problem of finite population inference in the context of longitudinal data with dropout and death. Death must be treated differently than non-response since post-death outcomes are truncated (and do not exist). Here we consider partly conditional inference and, as such, are interested in estimating the finite population mean given survival up to that time point, that is, the population partly conditional mean (PPCM). 
 
First we need some additional notation. For individual $i$ at time point $t=0,1,\ldots T$, denote the outcome variable and the vector of auxiliary information by $y_{it}$ and $\mathbf{x}_{it}$ respectively. The history of the time-varying variables are denoted with an overbar. For example, the outcome history for individual $i$ up to and including time point $t$ is denoted by $\bar{y}_{it}=\{y_{i0},y_{i1},\ldots,y_{it}\}$. Let $s_{it}$ denote survival, where $s_{it}=1$ if an individual is alive at time $t$ and 0 otherwise. Let $r_{it}$ be a response indicator, where $r_{it}=1$ if individual $i$ participates in the study at time $t$ and 0 otherwise. We assume monotone missingness, so if $r_{it}=0$, $r_{ik}=0$ for $k > t$, and of course similarly for $s_{it}$. Note that, $s_{it}$ is observed for all $i \in U$ and $r_{it}$ is observed if $i \in c$. Note that, the number of individuals in the population decreases over time due to deaths.

Initially we further assume the non-response mechanism to be missing at random conditional on being alive at time $t$ (MARS). 
That is, $p(y_{it} \mid \bar{y}_{it-1}, \bar{r}_{it-1},r_{it}=0, s_{it}=1, \bar{\mathbf{x}}_{it}) =$ $p(y_{it} \mid \bar{y}_{it-1},\bar{r}_{it}=1, s_{it}=1, \bar{\mathbf{x}}_{it})$. However, unlike previous works using MAR conditional on being alive (e.g. \cite{wen2018methods}), here time of death is known for all individuals in the population; hence, is does not need to be modeled.

The working model for making predictions under MARS becomes 
\begin{align}\label{WM_MARS}
\hat{y}_{it} = \int_{\bar{y}_{t-1}} \hat{m}_t(\bar{y}_{it-1},s_{it}=1, \bar{\mathbf{x}}_{it})\times \prod^{t-1}_{k=0} p(y_{ik}\mid y_{ik-1},s_{ik}=1,\bar{\mathbf{x}}_{ik}) d\bar{y}_{it-1}. 
\end{align}
Monte Carlo integration is used to integrate over the outcome history $\bar{y}_{t-1}$. To do this we factor the joint distribution of $\bar{y}_{t-1}$ into a sequence of one-dimensional conditional distributions. Population outcome data is then sequentially sampled from each of these, for all $i \in U$. Finally, predictions for $\hat{y}_{it}$ are obtained by plugging in $\bar{y}_{it-1}$ and $\bar{\mathbf{x}}_{it}$ given $s_{it}=1$ in $\hat{m}_t(\bar{y}_{it-1},s_{it}=1, \bar{\mathbf{x}}_{it})$.

The MB estimator of the PPCM at time $t$ is given by
$\mathrm{PPCM}^{MB}_t= \frac{1}{\sum_{i \in U} s_{it} }\sum_{i \in U} \hat{y}_{it}s_{it}$.
Note that, in context of the Betula study interest is not the $\mathrm{PPCM}_t$ at a specific test wave, but rather the age specific PPCM aggregated over test waves. That is, 
\begin{align*}
\mathrm{PPCM}^{MB}(age) = \frac{\sum_{i \in U_{age_t}} s_{it}}{\sum_{t=1}^T \sum_{i \in U_{age_t}} s_{it}} \sum_{t=1}^T \mathrm{PPCM}^{MB}_t(age),
\end{align*}
where $i \in U_{age_t}$ is the subset of individuals in age-cohort $age$ at test wave $t$, and $\mathrm{PPCM}^{MB}_t(age) = \frac{1}{\sum_{i \in U_{age_t}} s_{it} }\sum_{i \in U_{age_t}} \hat{y}_{it}s_{it}$.

\begin{center}
    \textit{3.3 Non-ignorable dropout among survivors and practice effects}
\end{center}
We introduce a set of sensitivity parameters to assess the impact of violations to the MARS assumption for the missingness mechanism. We additionally introduce a sensitivity parameter that account for practice effects (PEs). 
The general strategy is to model the observed data distribution, and to use priors on the sensitivity parameters to identify the full-data model.

Previous studies of the Betula data suggests that individuals who drop out have lower performance and steeper decline in memory (\cite{josefsson12}). Thus, we expect dropout to be missing not at random conditioning on survival (MNARS). To allow for deviations from the MARS assumption we introduce a parameter $\gamma_{it}$ to identify the expected outcome for dropouts among survivors. That is, for all $t>0$, $E(Y_{it} \mid \bar{y}_{it-1}, \bar{r}_{it-1},r_{it}=0, s_{it}=1, \bar{\mathbf{x}}_{it}) = E(Y_{it} - \gamma_{it} \mid \bar{y}_{it-1},\bar{r}_{it}=1, s_{it}=1, \bar{\mathbf{x}}_{it})$. Note that, after we condition on $\bar{r}_{it-1}$ we implicitly assume $y_{it}$ is conditionally independent of the wave at which the drop-out occurred. If $\gamma_{it}=0$ this implies a MARS assumption on the expectation for the outcome, and if $\gamma_{it}>0$ this implies a negative location shift in the outcome at the unobserved test wave. 

Practice effects (PEs) arise from increased familiarity with the assessment tools and may result in underestimating the decline in memory in longitudinal studies. Let $y_{it}^*$ denote the observed memory score (in contrast to $y_{it}$ which denotes an individual's actual memory function) and $\delta_{it}$ denote a sensitivity parameter. Then, for $t > 0$, $E(Y_{it} \mid \bar{y}_{it-1},\bar{r}_{it}=1, s_{it}=1, \bar{\mathbf{x}}_{it})=$ $E(Y_{it}^* - \delta_{it} \mid \bar{y}_{it-1},\bar{r}_{it}=1, s_{it}=1, \bar{\mathbf{x}}_{it})$, where $\delta_{it}>0$ implies an overestimated memory performance due to practice effects. Note that we assume no PEs at the initial testing. 

Our approach allows to explore sensitivity to the unverifiable assumptions by specifying informative priors for the sensitivity parameters $\gamma_{i1},\ldots,\gamma_{iT}$, and $\delta_{i1},\ldots,\delta_{iT}$. We specify triangular distributed priors conditioning on auxiliary variables, $\gamma_{it} \sim \mathrm{Tri}(A_t(\bar{\mathbf{x}}_{it}), 0, A_t(\bar{\mathbf{x}}_{it}))$ and $\delta_{it} \sim \mathrm{Tri}(0, B_t(\bar{\mathbf{x}}_{it}), B_t(\bar{\mathbf{x}}_{it}))$, where the three parameters of the Triangular distribution are the minimum, the mode and the maximum. 
We restrict the parameters of the Triangular distribution to a plausible range of values, reflecting the analysts' beliefs. Note that, using this distribution we are able to place more prior weight at one of the endpoints while still only needing to specify a lower- or a upper bound (and not an explicit variance parameter). In the Analysis of the Betula data we specify values of $A_t(\bar{\mathbf{x}}_{it})$ and $B_t(\bar{\mathbf{x}}_{it})$ in context of the study.

With the sensitivity parameters, identification of the PPCM based on the working model in (\ref{WM_MARS}), uses
\begin{align} \label{y_hat_MNAR}
\hat{y}_{it} = &\int_{\bar{y}_{t-1}^*} \sum_{\bar{r}_{it}} \Bigl\{ \left[ m_t(\bar{y}^*_{it-1},\bar{r}_{it},s_{it}=1, \bar{\mathbf{x}}_{it})  - \gamma_{it}I_{(r_{it}=0)} - \delta_{it}\right]   \times \nonumber \\
& \prod^{t-1}_{k=0} p(y_{ik}^* - \gamma_{ik}I_{(r_{ik}=0)} \mid y_{ik-1}^*,\bar{r}_{ik},s_{ik}=1,\bar{\mathbf{x}}_{ik}) \times \nonumber \\
& \qquad p(r_{ik} \mid  \bar{y}_{ik-1}^*, \bar{r}_{ik-1}, \bar{x}_{ik}, s_{ik}=1)\Bigr\} d\bar{y}_{ik-1}^*. 
\end{align}
An implication of (\ref{y_hat_MNAR}) is that Monte Carlo integration is implemented over both the outcome history, $\bar{y}_{t-1}$, and the response history, $\bar{r}_t$. The sensitivity parameters, $\gamma_t$ and $\delta_t$, must also be sampled for all $i \in U$. If $r_{ik}=0$, $\hat{y}_{ik}$ is shifted downwards by $\gamma_{ik}$ for $k=1,\ldots,t$. Predictions for $\hat{y}_{it}$ are similarly obtained by plugging in $\bar{y}_{it-1}$ and $\bar{\mathbf{x}}_{it}$ in the estimated model given $s_{it}=1$. Note that, for $t \geq 1$, $\hat{y}_{it}$ is further shifted downwards by $\delta_{it}$, thereby adjusting for PEs.
\begin{center}
\textsc{4 A semi-parametric approach for estimating the PPCM}
\end{center}
We propose a Bayesian semi-parametric modelling approach based on Bayesian Additive Regression Trees (BART; \cite{chipman2010bart}) for the working model in (\ref{y_hat_MNAR}) using the observed data and the sensitivity parameters.

\begin{center}
\textit{4.1 Semi-parametric estimation of the outcomes and dropout}
\end{center}
We specify BART models for the conditional distributions of the time varying variables $y_{it}$ and $r_{it}$. The distribution of the continuous outcome, $p(y_{it} \mid \bar{y}_{it-1},s \bar{\mathbf{x}}_{it})$ is specified as normal, $y_{it}\sim N\left(m_{t}(\bar{y}_{it-1}, \bar{\mathbf{x}}_{it}), \sigma_t^2\right)$, for the subset that satisfies $\bar{r}_{it}=1$ and $s_{it}=1$. 
The mean function at wave $t$, $m_{t}(\bar{y}_{it-1}, \bar{\mathbf{x}}_{it})$, is given by the sum of $k=1,\ldots,K_{y_t}$ regression trees, denoted by $g^k_{y_t}\left((\bar{y}_{it-1}, \bar{\mathbf{x}}_{it});T_{y_{t}}^k,M_{y_{t}}^k\right)$. Each regression tree consists of a set of decision rules (a tree structure), denoted by $T_{y_{t}}^k$, leading down to $b^k$ terminal node parameters, denoted by $M_{y_{t}}^k=(\rho_{y_{t}}^{k}(1),\ldots,\rho_{y_{t}}^{k}(b^k))$.
Hence, an individual's values for $(\bar{y}_{it-1}, \bar{\mathbf{x}}_{it})$ is linked to a single terminal node by following the decision rules for each tree, and is assigned the associated terminal node parameter. Similarly, the BART models for the binary response indicators $r_{it}$ are specified as probit models, $\pi_{it}(\bar{y}_{it-1}, \bar{\mathbf{x}}_{it})= \Phi \left(\sum_{k=1}^{K_{r_{t}}} g_{r_{t}}\left((\bar{y}_{it-1}, \bar{\mathbf{x}}_{it});T_{r_t}^k,M_{r_t}^k\right) \right)$, where $\Phi$ denotes the cumulative density function of the standard normal distribution and $\pi_{it}(\bar{y}_{it-1}, \bar{\mathbf{x}}_{it})$ is the probability of being observed at wave $t$ given $(\bar{y}_{it-1}, \bar{\mathbf{x}}_{it})$ for the subset that satisfies $\bar{r}_{it-1}=1$ and $s_{it}=1$. Note that, $r_{i0}=1$ and $s_{i0}=1$ for all individuals, and that $\pi_{it}=0$  if $r_{it-1}=0$.

\begin{center}
\textit{4.2 Algorithm}
\end{center}
The estimator of the PPCM as described in Section 3 can be computed using the algorithm in Table \ref{algorithm}. In practice, draws from the posterior distribution of the BART models are generated using Markov chain Monte Carlo (MCMC). The parameters of the models for $y_{it}$ and $r_{it}$ are assumed independent and thus their posteriors can be sampled simultaneously. We use the sparse Dirichlet splitting rule prior for BART (DART; \cite{linero2018DART}) to encourage parsimony, implemented in the R package \textit{BART} for continuous and binary responses. To simplify notation we denote the conditional probability $\pi_{it}(\bar{y}_{it-1},\bar{\mathbf{x}}_{it})$ by $\pi_{it}$ and the conditional expectation for the outcome at time $t$, $\mu_{it}(\bar{y}_{it-1} ,\bar{\mathbf{x}}_{it})$ by $\mu_{it}$.

\begin{center}
\textsc{5 Simulation study}
\end{center}
In this simulation study we compare five estimators for population inference. The sample and population size, the response rate, and the strength of association between $\hat{y}_{it}$ and $y_{it}$ (i.e. the $R^2$ estimated using BART), are chosen to mirror the Betula study. 

\begin{center}
    \textit{5.1 Data generating process}
\end{center}
We consider a finite population of size $N=10,000$, a sample of size $n=1,000$, and $1,000$ simulated datasets were generated. The auxiliary variables are generated independently as follows, $x_{1},x_{2} \sim Bernoulli(0.5)$ and $x_{3}, x_{4}, \ldots, x_{8} \sim Uniform(-1,1)$, where $x_{5} - x_{8}$ are uncorrelated with the outcome. We consider two time points ($t=0,1$). The response rate for $y_{i1}$ was set to approximately 75\%. Here, interest is in estimating the population mean at $t=1$, $\mu_U = N^{-1} \sum_{i\in U} y_{i1}$, for a continuous outcome variable at $t=1$. The approaches are compared in terms of bias, standard deviation (SD), mean squared error (MSE), and coverage of 95\% credible intervals.
We consider 5 true outcome models: 
\begin{enumerate}
\item \textit{Generalized linear additive models for the sample selection, response mechanism and outcomes}. Given the auxiliary variables $x_1 - x_8$, the outcome values for $i \in U$ were generated as
$y_{0i} = -1 - x_{1i} + x_{2i} + x_{3i} + x_{4i} + \varepsilon_{i1}$ and $y_{1i} = -1 - x_{1i} + x_{2i} + x_{3i} + x_{4i} - 0.3y_{0i} + \varepsilon_{i2}$, where $\varepsilon_i \sim N(0,1)$. The sample selection was generated from the following model
$\mathrm{logit}(\pi^c_i) = -2.67 - 0.4x_{1i} + 0.4x_{2i} + 0.4x_{3i} + 0.4x_{3i}$. In each selected sample, for $i \in c$ non-response to the study variable $y_1$ was generated from the following model
$\mathrm{logit}(\pi^r_i) = -2.7 + 1.2x_{1i} + 1.2x_{2i} + 1.2x_{3i} + 1.2x_{4i} - 1.2y_{0i}$.
\item \textit{Interaction and nonlinear dependencies for the response mechanism}. $y_{0i}$, $y_{1i}$ and $\pi^c_i$ were generated as for Scenario 1. The response mechanism was generated from the following model
$\mathrm{logit}(\pi^r_i) = -2.7 - x_{1i} + x_{2i} + x_{3i} + x_{4i} + y_{0i} + x_{3i}x_{4i} + x_{3i}x_{1i} + y_{0i}x_{1i}$.
\item \textit{Interactions, nonlinear dependencies and skew normal error terms}. $y_{0i}$, $\pi^r_i$ and $\pi^c_i$ were generated as for Scenario 2. $y_{1i}$ was generated according to
$y_{1i} = -0.87 - 0.4x_{3i} + 0.8x_{3i}^2 + 0.8x_{3i}^3 + 0.4x_{4i} + 0.8x_{1i} + 0.8x_{2i} + 0.4y_{0i} - 0.4x_{1i}y_{0i} + \varepsilon_{it}$, where $\varepsilon_{it}$ were generated from the skew normal distribution, such that $\varepsilon_{it} \sim SN(-1.6\times\frac{5}{\sqrt{1+5^2}}\times\sqrt{\frac{2}{\pi}},1.6,5)$ for t=0,1, i.e. a right skewed variable with 0 mean, a variance of 1 and a skewness of 1.3. 
\item \textit{Interactions, nonlinear dependencies and practice effects}. $y_{0i}$, $\pi^r_i$ and $\pi^c_i$ were generated as for Scenario 3. For the outcome $y_{1i}$ we added a moderate PE of 0.1 to Scenario 3.
\item \textit{Interactions, nonlinear dependencies and deaths}.
While $y_{0i}$, $y_{1i}$, $\pi^r_i$, and $\pi^c_i$ were generated as for Scenario 3, survival at $t=1$, $s_{i1}$, was generated from $\mathrm{logit}(\pi^s_i) =$ $1.7 + 0.35x_{1i} + 0.35x_{2i} + 0.35x_{3i} + 0.35x_{4i}$, for a overall death rate of 12\%.
\end{enumerate}

\begin{center}
    \textit{5.2 Estimators for population inference}
\end{center}
We compare five estimators for population inference. Our semi-parametric model based approach (MB-sp), was implemented as described in Section 4 and Table \ref{algorithm}, but fixing the sensitivity parameters to 0. However, for the fourth scenario we used the PE sensitivity parameter and specified a triangular prior reflecting practice effects, $\delta_{i1} \sim \mathrm{Tri}(0,b,b)$ with $b=0.05,0.1,0.15$. Note that for the other approaches the sensitivity parameters were set to 0 for all scenarios.
Details of the other four estimators (MB-lm, HT, MRP, and GREG) for longitudinal data are given in Appendix A of the Supplementary materials. Briefly, MB-lm, is a parametric version of the MB-sp, specifying the working models as Bayesian additive linear regression models instead of using DART. Default uninformative priors were used. HT is an extension of the classic Horvitz-Thompson weighting estimator, where the inclusion weights were replaced by longitudinal \textit{probability of participation weights}. The general regression estimator (GREG), combines prediction and longitudinal weighting. For GREG, the weights were computed as for HT and the prediction models were estimated using additive linear regression models. MRP is an extension of multilevel regression and poststratification (\cite{gelman1997poststratification}), where the binary covariates and the outcome at $t=0$ were added to the model as fixed effects and the six continuous variables were first categorized into quartiles and added to the model as random effects. 

The inclusion weights used in HT and GREG must be estimated using cell weight adjustment. To avoid sparse cells only $x_1 -x_4$ were considered for computing the cell adjustment weights, hence, the uncorrelated variables $x_5 - x_8$ were omitted. The continuous variables $x_3$ and $x_4$ were first categorized into tertiles. Every unique combination of the categorized variables constituted an adjustment cell. Sparse cells, $n_j< 20$, were identified and combined with their nearest, non-sparse neighbor, i.e. the cell that has the most similar combination of auxiliary variables. Weights larger than 30 were trimmed.

The HT and GREG estimators and their 95\% confidence intervals were estimated using the \textit{mase} package in R, the working models for MB-lm and MRP were fitted using the \textit{MCMCpack} package and \textit{MCMCglmm} package in R (\cite{R2018}). MCMC convergence and mixing was monitored using trace plots.

\begin{center}
    \textit{5.3 Simulation results}
\end{center}
We present the bias, empirical standard deviation (SD; calculated as the standard deviation of the parameter estimates), mean-squared error (MSE), and coverage probabilities for the 95\% credible intervals (CP) from the 1,000 simulations in Table \ref{Table_sim_results}. The results show that inferences using the sample are highly biased, have the highest MSEs and the lowest coverage probabilities under all four scenarios ($\leq 3.7\%$). The weighting approach (HT) performs poorly with large bias, SDs and MSEs, and lowest CPs compared to the other 4 approaches across Scenario 1-4.

Scenario 1 and 2 gives similar findings for MB-sp, MB-lm, MRP, and GREG. That is, all approaches are unbiased, or nearly unbiased, and have similar SDs and MSEs. MB-sp and MB-lm have CPs above or close to $95\%$, while MRP and GREG have CPs just below $90\%$. Under Scenario 3 and 4, MB-lm, MRP and GREG are severely biased and their CPs are low (Scenario 3: $6.9\%$ - $42.0\%$ and Scenario 4: $0.1\%$ - $2.1\%$). MRP has highest biases and MSEs, and lowest CPs. In contrast, under Scenario 3 our MB-sp approach is nearly unbiased, has the lowest MSE, and a CP of $91.6\%$. Under Scenario 4 (where the true PE was 0.1), our MB-sp approach revealed a bias of 0.118 (SD=0.046) and a CP of $23.3\%$. Which was nevertheless best among all the approaches. For the analyses adjusting for practice effects, when the upper bound (and mode) of the Triangular prior was set to $0.05$, the bias was 0.087 (SD=0.048). If the upper bound was set to the true PE, $0.1$, the bias was further reduced to 0.045 (SD=0.043). Finally, if the upper bound was set to $0.15$ (with a mean equal to the true PE) the bias was 0.019 (SD = 0.047) and the CP was $90.1\%$, which is comparable to Scenario 3. The CPs were, $50.0\%$ and $85.2\%$ for the upper bounds of $0.05$ and $0.1$ respectively. These results demonstrate the importance of the PE adjustment. For Scenario 5, adjusting for truncation due to death, we compared, for simplicity of implementation, results for the three model based approaches. MB-sp, MB-lm and MRP revealed results comparable to Scenario 3, i.e. MB-sp: bias$=0.023$ $(SD=0.047)$ and $CP=92.3\%$, MRP: bias$=0.195$ $(SD=0.06)$ and $CP=2.4\%$, and MB-lm: bias$=0.115$ $(SD=0.050)$ and $CP=32.6\%$.

\begin{center}
\textsc{6 Analysis of the Betula data}
\end{center}
We applied the proposed semi-parametric approach for estimating the $\mathrm{PPCM}(age)$ to the Betula data. Here, interest is in estimating the average memory performance across the adult lifespan among non-demented individuals given survival up to a specific age. Note, that both death and dementia are considered to be sources of truncation in the analysis. We consider population inference in the context of longitudinal cohort data in the presence of practice effects, non-ignorable dropout and death. Longitudinal auxiliary data for both the sample and target population is available for four waves of data collection. These include the baseline variables age, sex, having children (Y/N), and highest level of education, and additionally, several income variables, benefits received from the government, and marital status were treated as time-varying. Details of baseline characteristics are found in Table \ref{Descriptives}. Details of the International Classification of Diseases (ICD) codes are given in Appendix B of the Supplementary materials.

\begin{center}
    \textit{6.1 Sensitivity parameters}
\end{center}
We explore sensitivity to default assumptions by specifying informative priors for the sensitivity parameters. Since there is generally little information about the distributional form for the SPs, the prior distributions is often chosen by the analyst reflecting prior beliefs about the departures from the default assumptions.

We assume triangular priors for $\delta_{it}$, reflecting the improvement in memory performance that occur at repeated testings, i.e. practice effects (PEs). The priors were specified as $\delta_{i1} \sim \mathrm{Tri}(0, U_{\delta_{1}}(\mathrm{a}_{i1}), U_{\delta_{1}}(\mathrm{a}_{i1}))$ and 
$\delta_{i2},\delta_{i3} \sim \mathrm{Tri}(0, U_{\delta_{2}}(\mathrm{a}_{i2}), U_{\delta_{2}}(\mathrm{a}_{i2}))$, where $\mathrm{A}_{it}=\mathrm{a}_{it}$ is the age of individual $i$ at wave $t$.  
The upper bounds are given by; $U_{\delta_{1}}(\mathrm{a}_{i1}) = 4.8-0.1\times\mathrm{a}_{i1} + 5.2*10^{-4}\times\mathrm{a}_{i1}^2$ and $U_{\delta_{2}}(\mathrm{a}_{i2}) = 11.0-0.3\times\mathrm{a}_{i2} + 1.9\times10^{-3}\times\mathrm{a}_{i2}^2$.
To obtain these we considered two comparable samples (S1 and S3) from the Betula study. The estimated practice effects were derived as the mean differences in memory performance between participants taking the test for the 2nd/3rd time (sample S1) versus those taking the test for the 1st/2nd time (sample S3), while adjusting for age and age squared. Hence, $U_{\delta_{t+1}} = E(Y_{it+1} | \mathrm{A}_{it+1}=\mathrm{a}_{it+1}, S1, R_{it+2}=1)-$ $E(Y_{jt} | \mathrm{A}_{jt}=\mathrm{a}_{jt+1}, S3, R_{jt+1}=1)$, for $t=0,1$, and $j=1,\ldots,N_{S3}$ refers to individuals in sample S3. By conditioning on $R_{it+1}$ we avoid non-response bias. Note that, practice effects were larger for younger participants and at $t=1$ compared to $t=0$.
We similarly specify a triangular prior for $\gamma_{it} \sim \mathrm{Tri}(L_{\gamma_t}(\mathrm{a}_{it}), 0, L_{\gamma_t}(\mathrm{a}_{it}))$, reflecting a decline in memory after dropping out of the study. The bounds are given by; $L_{\gamma_t}(\mathrm{a}_{it}) =-8.0 - 0.3 \times \mathrm{a}_{it} +3.9\times10^{-3}\times\mathrm{a}_{it}^2$, obtained from estimating, $E(Y_{i1}-Y_{i0} | \mathrm{A}_{i1}=\mathrm{a}_{i1}, S1, R_{i1}=1)$, i.e. the change in memory performance between the first and second wave for responders while adjusting for age and age squared. Note that, this imply that $\gamma_{it}$ is the same for all $t$ and that older participants decline more quickly than younger participants after drop out.

\begin{center}
    \textit{6.2 Results}
\end{center}
We estimated the PPCM using our proposed Bayesian semi-parametric approach with the sensitivity parameters. For each chain the first 1000 iterations were discarded as burn-in, and a total of 1000 posterior samples of the PPCM were obtained. Convergence of the posterior samples was monitored using trace plots.

In the main analysis the priors for the SPs were specified as described in the previous section. For the sensitivity analysis we compare the results for different values of the SPs. Specifically, the SPs were set to
i) $\gamma_{it}=0$ while $\delta_{it}$ was specified as in the main analysis (MARS and PE adjustment), 
ii) $\delta_{it}=0$ for $t=2,3,4$, while $\gamma_{it}$ was specified as in the main analysis (MNARS and no PE adjustment),
iii) the SPs were specified as twice as large as in the main analysis ($2 \times \delta_{it}$ for $t=2,3,4$, and similarly, $2 \times \gamma_{it}$), and
iv) $\gamma_{it}=0$ and $\delta_{it}=0$ for $t=2,3,4$ (MARS and no PE adjustment).
We moreover compared the results from our mortal-cohort analysis to an analysis assuming an immortal cohort (\cite{wen2018methods}), treating both death and dropout as MAR ($\gamma_{it}=0$); hence, deceased participants are implicitly included after death. In this scenario only baseline auxiliary variables were considered since $x_{it}$ is missing when $s_{it}=0$ and $\delta_{it}=0$ for all t (no PE adjustment).

\begin{figure}[hbt!]
\includegraphics[width=1.0\textwidth]{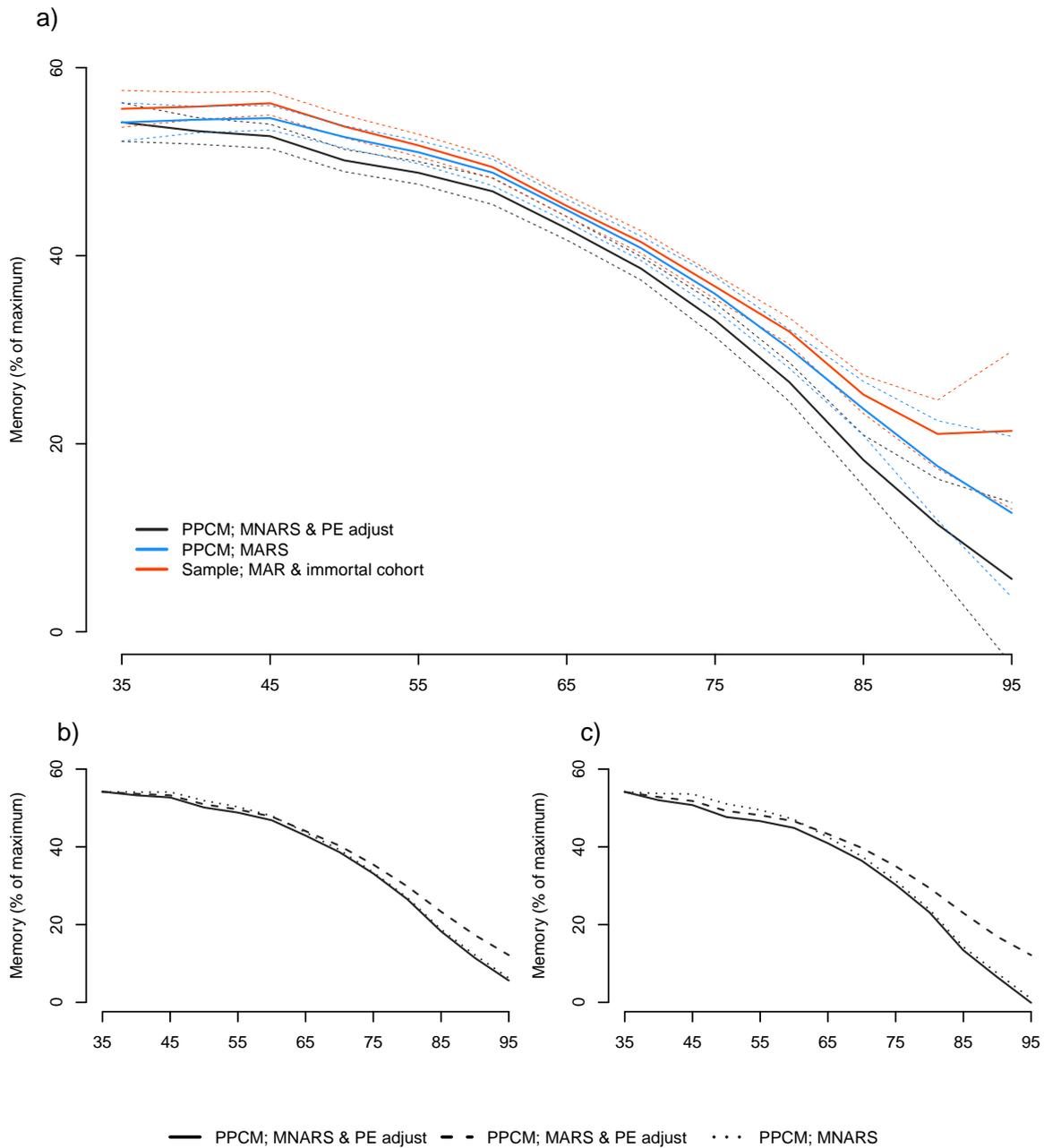} 
\caption{Estimating lifespan trajectories of memory for the Betula data using our MB-sp approach. Panel a) \textit{PPCM, MNARS \& PE}: the main analysis; \textit{PPCM, MARS}: the sensitivity parameters were all set to 0; \textit{Sample, MAR \& immortal cohort}: sample inference assuming MAR missingness for both dropouts and deaths and no practice effects. Panel b) and c) shows a sensitivity analysis for k=1 and k=2 respectively, comparing the main analysis (\textit{PPCM, MNARS \& PE}) to an analysis where $\gamma_i=0$ and $\delta_{it}$ were specified as for the main analysis (\textit{PPCM, MARS \& PE}), and to an analysis where $\delta_{it}=0$, while $\gamma_i$ was specified as as for the main analysis (\textit{PPCM, MNARS}).} 
\label{Fig_Betula} 
\end{figure}

The results from the main analysis and the various sensitivity analyses are presented in Figure \ref{Fig_Betula}. The corresponding 95\% credible intervals are also plotted for the main analysis, the immortal-cohort analysis and for the analysis assuming MARS and no PE adjustment (Figure \ref{Fig_Betula}a). For the main analysis, the age-specific $\mathrm{PPCM}$ revealed an initial decline in memory performance between the ages 35 to 65, and accelerated decline after the age of 65. Assuming an immortal cohort, no PEs and MAR non-response (both for dropout and death) resulted in a significantly higher estimated memory performance across adulthood, with slightly greater decline at younger ages and less decline in memory after the age of 65. 
An analysis assuming MARS missingness and not adjusting for practice effects resulted in a significantly higher estimated memory performance between the ages $40 - 90$ compared to the main analysis. 

It is apparent in Figure \ref{Fig_Betula}b and \ref{Fig_Betula}c that adjusting for both PEs and MNARS dropout resulted in lower estimated memory performance across adulthood, although the discrepancy varied in magnitude with respect to age. When the SPs were specified as twice as large as in the main analysis (\ref{Fig_Betula}c), we see an increased reduction in memory performance, this is more pronounced at older ages. As expected, comparing a PE adjusted analysis to an analysis without PE adjustment (assuming MNARS dropout) revealed lower memory performance for the PE adjusted analysis, although the discrepancy was more pronounced at younger ages. Comparing results from an analysis assuming MNARS and MARS while adjusting for PEs, also revealed lower memory performance, and the difference was more pronounced at older ages. 

We compare the results for the Betula data using our approach (MB-sp) with the four other estimators for population inference, MB-lm, HT, MRP, and GREG. The sensitivity parameters, $\gamma_{it}$ and $\delta_{it}$ for $t=1,2,3$, were for simplicity all set to 0 when estimating the PPCM, i.e assuming MARS missingness and no PEs.

Details of the estimation procedures for MB-lm, HT, MRP, and GREG are described in Section 5.2 and in Appendix A of the Supplementary materials. However, some adjustments were made. For MRP, the continuous variables were first categorized into quartiles plus an additional category if the variable was 0 (added to the model as random effects). Due to excessive number of cells with a small or zero sample size we only used a subset of the baseline auxiliary variables for computing the cell adjustment weights $\pi_{i}$ in HT and GREG. These were age, sex, level of education, widowhood, and history of cardiovascular disease.

\begin{figure}[hbt!]
\includegraphics[width=1.0\textwidth]{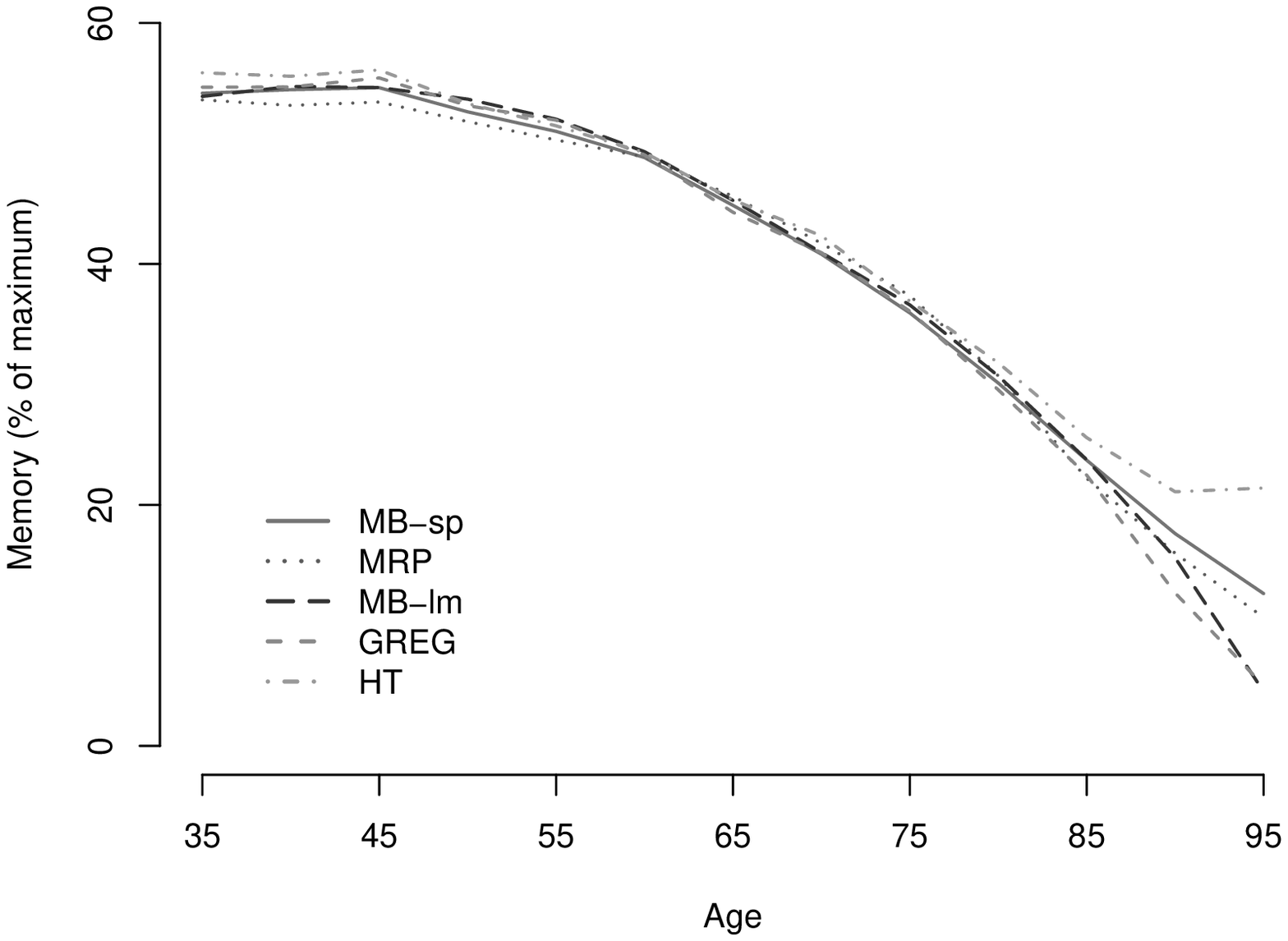} 
\caption{Analysis of the Betula data. Results from estimating the population partly conditional mean using 1) MB-sp: our semi-parametric model-based approach using DART, 2) MB-lm: a parametric model based approach using additive linear regression, 3) MRP: Multilevel regression and poststratification, 4) GREG: general regression estimator, and, 5) HT: the Horvitz-Thompson estimator. The sensitivity parameters $\gamma_i, \delta_{i2}, \delta_{i3}$ and $\delta_{i4}$ were all set to 0.} 
\label{Fig_Betula_comp_other_methods} 
\end{figure}

Results are found in Figure \ref{Fig_Betula_comp_other_methods}. Compared to our semi-parametric approach (MB-sp), the HT estimator overestimated memory performance across adulthood, and the three other parametric approaches revealed accelerated decline in memory performance for individuals 70 - 95 years old. The large discrepancy in memory performance for HT, and for GREG at older ages, is likely a result from using only a subset of the auxiliary variables, dichotomizing continuous variables and collapsing zero or spars cells, since this may introduce bias in the weights. Given the findings of the simulation study, we are inclined to believe the discrepancies for the parametric approaches (MB-lm and MRP) from our semi-parametric approach are likely due to model-misspecification. Note that, MRP gives results most similar to our semi-parametric approach. 

\begin{center}
\textsc{7 Conclusions}
\end{center}
This article proposes a Bayesian semi-parametric predictive estimator for estimating the population partly conditional mean, when a large set of longitudinal auxiliary variables is known for all units in the target population. A key feature is the flexible modeling approach that effectively addresses nonlinearity and complex interactions. Additionally, BART (using the sparse Dirichlet splitting rule prior) demonstrated excellent predictive performance when irrelevant regressors were added, diminishing the need to carry out formal variable or model selection.

Our study is motivated by the fact that it is becoming increasingly difficult to recruit study participants, which may severely distort the ability to generalize study findings. The increased availability of microdata covering the population in many countries however, makes post-sampling adjustments an attractive tool. 
Although weighting is the most popular technique, a large set of auxiliary (possibly continuous) variables makes cell weight adjustment difficult.
In this setting model based approaches are more attractive, but put stronger requirements on correct model specification. As expected, the results of the simulation study showed that the weighting approach (HT) performed poorly across a wide range of scenarios, despite a simplified scenario where uncorrelated variables were excluded. This is likely a result from collapsing sparse cells and dichotomizing the continuous auxiliary variables, thereby introducing bias in the weights. In contrast, the model based approaches and GREG all performed well under correct specification of the outcome model, although, our semi-parametric method was the only approach who gave unbiased results for the more realistic scenario with unknown nonlinearity and interactions. Furthermore, under the scenario with practice effects our approach performed relatively well compared to the other approaches, showing the importance of adjusting for practice effects.

The goal of the empirical study was to estimate lifespan trajectories in memory for a target population. The results revealed an initial decline in memory performance between the ages 35 to 65, in contrast to previous studies showing stable performance up to the age of 60 (\cite{ronnlund05}, \cite{gorbach2017longitudinal}). Furthermore, the standard approach for estimation in previous literature that assumes an immortal cohort, no PEs and MAR non-response revealed significantly higher memory performance across the adult lifespan compared to our approach. This suggests that in previous studies the magnitude of memory performance across adulthood is likely overestimated while the rate of change is likely underestimated, especially at older ages. This is due to both selective study enrollment and attrition.

Our approach allows for Bayesian inference under MNAR missingness and truncation by death, as well as the ability to characterize uncertainty about practice effects. This was accomplished by introducing sensitivity parameters (SPs) that incorporated prior beliefs. A strength of the current approach is that inference with SPs and a mortal cohort is relatively easy to implement and communicate to non-statisticians. However, specifying an appropriate prior distribution can sometimes be difficult; a alternative approach could be a tipping point analysis (\cite{yan2009missing}). In a tipping point analysis subject matter experts can discuss whether the tipping point for the SPs are plausible, which may aid in making judgment based on study findings.  

\begin{center}
\textsc{Supplementary online material}
\end{center}
In Appendix A of the Supplementary materials we provide details for the alternative estimators for the PPCM. Classification of Diseases (ICD) codes considered are given in Appendix B. R code for the main analysis of the Betula data and the simulation study using our approach is made available upon request to the corresponding author.

\begin{center}
\textsc{Acknowledgments}
\end{center}
This work was supported by The Swedish Foundation for Humanities and Social Sciences [P17-0196:1 to M.J.], and National Institutes of Health [R01 CA 183854 to M.J.D., R01 GM 112327 to M.J.D.]. We would like to thank Prof. Rolf Adolfsson and project coordinator Annelie Nordin Adolfsson, Department of Clinical Sciences at Umeå University, for providing data from the Betula project for analyses. The Betula Project is supported by Knut and Alice Wallenberg foundation and the Swedish Research Council (K2010-61X-21446-01).

\begin{center}
\textsc{References}
\end{center}
\printbibliography[heading=none]

\newpage
\subfile{Supplementary}

\newpage
\begin{center}
\textsc{Tables and Figure}
\end{center}
\begin{table}[hbt!]
\caption{Algorithm for estimation of the PPCM as described in Sections 3 and 4.} 
\label{algorithm}
\begin{tabular}{ l l}
\hline
1. & \textit{Models for the outcomes and response mechanisms}:\\
&For $t=1,\ldots,T$, sample from the observed data posteriors for the parameters of the \\
&conditional distribution of $y_t$ and $r_t$ using DART. \\
2. & \textit{Sensitivity parameters}: \\
&For all $i\in U$ and $t=1,\ldots,T$, sample one set from the prior distributions for $\gamma_{it}$ \\
& and $\delta_{it}$. \\
3. & \textit{Compute predicted means}: \\
& For all $i\in U$ and $t=1,\ldots,T$, compute $\hat{y}_{it}$ in [\ref{y_hat_MNAR}] as follows:\\
&Sequentially (in $t$) sample response and outcome data from $r_{it}^* \sim \mathrm{Ber}\big(\pi_{it}\big)$ and \\
&$y_{it-1}^* = \hat{m}_{t-1}(\bar{y}^*_{it-2},\bar{r}^*_{it-1},s_{it-1}=1, \bar{\mathbf{x}}_{it-1}) + \gamma_{t-1}I_{(\hat{r}_{it-1}=0)} + \varepsilon_{it}^*$. 
The conditional \\
&probability $\pi_{it}$ conditions on $\bar{\mathbf{x}}_{it}$, $\bar{y}^*_{it-1}$, $\bar{r}^*_{it-1}$. For all $s_{it}=1$, compute \\
&$\hat{y}_{it}= \hat{m}_t(\bar{y}^*_{it-1},\bar{r}^*_{it},s_{it}=1, \bar{\mathbf{x}}_{it})  - \gamma_{it}I_{(r^*_{it}=0)} - \delta_{it}$.\\
4. & \textit{Compute the PPCM}: \\
&Compute one posterior sample of $\mathrm{PPCM} = \frac{1}{\sum_{t=1}^T \sum_{i \in U} s_{it}} \sum_{t=1}^T \sum_{i \in U} \hat{y}_{it}s_{it}$. \\ 
5. & Repeat step 2 - 4 for each posterior sample.\\
\hline
\end{tabular}
\end{table}

\begin{table}[hbt!]
\caption{Results from 1,000 simulations with a finite population of size 10,000 and a sample size of 1,000. Bias, empirical standard deviation (SD; calculated as the standard deviation of the parameter estimates), mean-squared error (MSE), and coverage probabilities in \% for the 95\% credible intervals (CP). Bias and SD are multiplied by 100 for ease of presentation. \textit{Abbreviations}: Sample: inference using the sample. MB-sp: a semi-parametric model based approach using DART; MB-lm: a parametric model based approach using additive linear regression; MRP: Multilevel regression and poststratification; GREG: general regression estimator; HT: the Horvitz-Thompson estimator.}
\centering
\begin{tabular}{lcccccccc}
  \hline
Method & \multicolumn{4}{c}{Scenario 1} & \multicolumn{4}{c}{Scenario 2} \\
  \hline
  & Bias & SD & MSE & CP & Bias & SD & MSE & CP \\ 
Sample & 18.8 & 5.3 & 3.8 & 3.7 & 18.8 & 5.1 & 3.8 & 3.1 \\ 
  MB-sp & 0.5 & 4.2 & 0.2 & 93.6 & 0.4 & 4.3 & 0.2 & 94.2 \\ 
  MB-lm & 0.2 & 3.9 & 0.2 & 95.2 & -0.1 & 4.2 & 0.2 & 94.5 \\ 
  MRP & 1.6 & 4.3 & 0.2 & 89.6 & 1.6 & 4.6 & 0.2 & 88.9 \\ 
  GREG & 0.1 & 4.1 & 0.2 & 89.6 & -0.1 & 4.2 & 0.2 & 89.5 \\ 
  HT & 4.8 & 4.7 & 0.4 & 86.7 & 7.2 & 4.7 & 0.7 & 73.9 \\ 
  \hline
 & \multicolumn{4}{c}{Scenario 3} & \multicolumn{4}{c}{Scenario 4}\\
  \hline
  & Bias & SD & MSE & CP & Bias & SD & MSE & CP \\ 
Sample & 41.1 & 5.6 & 17.2 & 0.0 & 51.0 & 5.5 & 26.3 & 0.0 \\ 
  MB-sp & 1.5 & 4.9 & 0.3 & 91.6 & 11.8 & 4.6 & 2.6 & 23.3 \\ 
  MB-lm & 11.8 & 5.1 & 1.7 & 33.0 & 21.7 & 5.0 & 5.0 & 0.9 \\ 
  MRP & 19.7 & 5.4 & 4.2 & 1.9 & 29.6 & 5.7 & 9.1 & 0.1 \\ 
  GREG & 9.1 & 5.4 & 1.1 & 45.5 & 18.6 & 5.2 & 3.7 & 2.1 \\ 
  HT & 21.2 & 6.4 & 4.9 & 6.9 & 30.9 & 6.3 & 9.9 & 0.3 \\ 
  \hline
\end{tabular}
\label{Table_sim_results}
\end{table}

\begin{table}[hbt!]
\caption{Baseline descriptive statistics for the Betula sample and the target population. Mean (standard deviation) for continuous variables and N (\%) for categorical variables. \textit{Abbreviations}: CVD: cardiovascular disease. * amounts in Swedish kronor rounded to the nearest thousand.}
\centering
\begin{tabular}{lcc}
\hline
& Sample & Population \\ 
\hline
N & 1000 & 9203 \\
Age & 57.5 (14.4) & 52.0 (13.3) \\ 
Male & 470 (47.0) & 4431 (48.1) \\ 
Having children & 856 (85.6) & 7952 (86.4) \\ 
$\leq 9$ years of education & 308 (30.8) & 2505 (27.2) \\ 
$10 - 12$ years of education & 352 (35.4) & 3288 (35.7) \\ 
$> 12$ years of education & 240 (24.2) & 2516 (27.3) \\ 
Education level unknown & 100 (10.1) & 894 (9.7) \\ 
Married & 628 (63.2) & 5960 (64.8) \\ 
Widow & 149 (14.9) & 715 (7.8) \\ 
Disposable income* & 1021.3 (412.8) & 1058.6 (490.5) \\ 
Earned income* & 868.9 (936.2) & 1045.2 (952.6) \\ 
Retirement income* & 275.1 (404.3) & 165.4 (335.0) \\ 
Early retirement income* & 37.2 (172.6) & 29.6 (151.1) \\ 
Unemployment benefits* & 4.4 (48.3) & 9.1 (64.5) \\ 
Health benefits* & 68.7 (220.0) & 86.7 (227.8) \\ 
Diagnosed with diabetes prior 1990 & 14 (1.4) & 185 (2.0) \\ 
Diagnosed with CVD prior 1990  & 97 (9.7) & 593 (6.4) \\ 
\hline
\end{tabular}
\label{Descriptives}
\end{table}

\end{document}

%% file: Supplementary.tex
\thispagestyle{plain}
\begin{center}
    {\large
    \textbf{Supplementary material - A Bayesian semi-parametric approach for inference on the population partly conditional mean from longitudinal data with dropout}}
        
\vspace{0.25cm}
\textsc{Josefsson, Maria$^{1,2,3,*}$, Daniels, Michael J.$^4$, \& Pudas, Sara$^{3,5}$}
\end{center}

\begin{center}{\footnotesize
$^1$Department of Statistics, USBE, Ume{\aa} University, Sweden.

$^2$Centre for Demographic \& Ageing Research, Ume{\aa} University, Sweden.

$^3$Ume{\aa} Center for Functional Brain Imaging, Ume{\aa} University, Sweden.

$^4$Department of Statistics, University of Florida, USA.

$^5$Department of Integrative Medical Biology, Ume{\aa} University, Sweden.

$^*$Corresponding author, E-mail: maria.josefsson@umu.se}
\end{center}

\begin{center}
\textsc{Appendix A, Other strategies for estimating the PPCM}
\end{center}
In the simulation study and the empirical data example we compare our proposed approach to four other estimators for estimating the PPCM: MB-lm, MRP, HT and GREG. 

The first estimator (MB-lm) is a parametric version of our semi-parametric approach, where the working models are specified as Bayesian additive linear regression models instead of using DART. 

The second approach is multilevel regression and poststratification (MRP). The MRP specifies a multilevel model for the outcome at time $t$, using random (or modeled) effects for some of the predictors. That is, the working model is replaced by the following multilevel regression model 
$$y_{it}= \alpha + \sum_{m=0}^{t-1} \beta_m y_{im} + \sum_{m=0}^{t} \sum_{k=1}^K b_{x_{mk}} + \varepsilon_{it},$$
where $\alpha$ is the fixed intercept, $\beta_0, \ldots, \beta_{t-1}$ are the fixed effects for the outcome history, and $b_{x_{1}j_{x_{k}}}, \ldots,b_{x_{K}j_{x_{k}}}$ are the random effects for the $j_{x_{k}}$th category of the auxiliary variable $x_k$, where  $k=1,\ldots,K$, and $j_{x_{k}}=1,\ldots,J_{x_{k}}$. All random effects are modeled using independent normal distributions, such that $b_{x_{k}} \sim N(0, \sigma^2_{x_{k}})$, $k=1, \ldots, K$, and, $\varepsilon_{it}\sim N(0,\sigma^2_{MRP_t})$. In the second step, the multilevel model is used for making predictions $\hat{y}_{it}$ from $\bar{y}_{it-1}$ and $\bar{\mathbf{x}}_{it}$, for all $i \in U$ given $\bar{s}_{it}=1$, and is (similarly to our approach) obtained by integrating over the outcome history $\bar{y}_{t-1}$. The poststratification estimate for the population mean is given by $\mathrm{PPCM}^{MRP}_t= \frac{1}{\sum_{i \in U} s_{it} }\sum_{i \in U} \hat{y}_{it}s_{it}$.

The third approach is an extension of the classic Horvitz-Thompson weighting estimator (HT), $\hat{\mu}^{HT} = \frac{1}{N} \sum_{i \in c} \frac{y_i}{\pi_i}$. In a longitudinal setting with MAR missingness and death, the HT estimator is obtained by replacing $\pi_i$ with the probability of participation at time $t$ given survival at that time point, here denoted by $w_{it}$. The $\mathrm{PPCM}_t$ is given by,
$$\mathrm{PPCM}^{HT}_t= \frac{1}{\sum_{i \in U} s_{it}} \sum_{i \in c}\frac{y_{it}}{w_{it}}\sum_{i \in c} r_{it}s_{it},$$
where $$w_{it} = \pi_{i} \prod_{k=0}^t\Pr(r_{ik}=1 \mid \bar{y}_{ik-1}, \bar{r}_{ik-1}, \bar{x}_{ik}, \bar{s}_{ik}=1)$$ 
for all $i \in c$. The response mechanism, the second term on the right hand side, can be estimated from data using e.g. a logistic regression model. In the empirical study using the Betula data, two separate datasets are used for finite population inference and the participation probability is not known. Then $\pi_i$ must be estimated from data using cell weight adjustment. Cell weight adjustment classifies the sample and population into distinct post-stratification cells based on the auxiliary variables recorded for both groups. The sampled participants in cell $j$ are weighted by the inverse of the sampled rate in cell $j$. That is, for individual $i \in j$, $\hat{\pi}_{i} = n_{j}/N_{j}$, where $n_{j}$ is the number of individuals in cell $j$ in the sample and $N_{j}$ is the number of individuals in cell $j$ in the population. However, difficulties may arise in this setting. For example, continuous variables have to be dichotomized, and further, with a large set of auxiliary variables, the cell sample sizes can be small. This may result in biased and unstable estimates. To overcome the latter, weight trimming and cell collapsing is recommended. 

The fourth approach is a dual-modelling strategy that combines prediction and weighting. The general regression estimator (GREG) at time $t$ is given by, 
$$\mathrm{PPCM}^{GREG}_t = \frac{1}{\sum_{i \in U} s_{it}} \left[ \sum_{i \in U} \hat{y}_{it} + \sum_{i \in c}r_{it}s_{it}\left(\frac{y_{it} - \hat{y}_{it}}{\hat{w}_{it}}\right) \sum_{i \in c} r_{it}s_{it}  \right].$$ 
The approach require both a model for the outcome and the participation mechanism, and is double robust in the sense that it remains consistent if either one of the models are correctly specified.

\begin{center}
\textsc{Appendix B, ICD codes}
\end{center}
The following ICD codes extracted from the in-patient and cause of death registers were considered:
Dementia, ICD8, ICD-9: 290 and ICD-10: F00 - F01, F03. CVD, ICD8, ICD-9: 410, 411, 412, 413, 414, 428, 430, 431, 432, 433, 434, 	435, 436, 437, 438, 440, 441, 442, 443, 444, 445, 446, 	447, 448, and ICD-10:	I21, I60 - I63. Diabetes, ICD8, ICD-9:	250, and ICD-10:	E10 - E14. Depression, ICD8, ICD-9:	296, and ICD-10: F32 - F33. Alcohol substance abuse, ICD8, ICD-9: 291, and ICD-10:	F10.